\documentclass[12pt]{article}
\usepackage{epsfig}
\def\be{\begin{equation}}
\def\ee{\end{equation}}
\def\bea{\begin{eqnarray}}
\def\eea{\end{eqnarray}}
\usepackage{graphicx}

\catcode`\@=11
\def\lsim{\mathrel{\mathpalette\@versim<}}
\def\gsim{\mathrel{\mathpalette\@versim>}}
\def\@versim#1#2{\vcenter{\offinterlineskip
\ialign{$\m@th#1\hfil##\hfil$\crcr#2\crcr\sim\crcr } }}
\catcode`\@=12
\usepackage{axodraw}

\catcode`@=12
\begin{document}
\thispagestyle{empty}
\begin{flushright}
UCRHEP-T524\\
October 2012\
\end{flushright}
\vspace{0.6in}
\begin{center}
{\LARGE \bf Heptagonic Symmetry for\\[5pt] Quarks and Leptons\\}
\vspace{1.5in}
{\bf Subhaditya Bhattacharya, Ernest Ma, Alexander Natale,\\ 
and Daniel Wegman\\}
\vspace{0.5in}
{\sl Department of Physics and Astronomy, University of California,\\
Riverside, California 92521, USA\\}
\end{center}
\vspace{1.0in}
\begin{abstract}\
The non-Abelian discrete symmetry $D_7$ of the heptagon is successfully 
applied to both quark and lepton mass matrices, including $CP$ violation.
\end{abstract}

\newpage
\baselineskip 24pt

\section{Introduction}
The structure of quark and lepton mass matrices has been under theoretical 
study for many years.  Whereas the 6 quark masses and the 3 mixing angles 
and 1 $CP$ violating phase in the quark sector are now measured with some 
precision, the lepton sector is still missing some crucial information. 
Recently, the neutrino mixing angle $\theta_{13}$ has been measured by the 
Daya Bay~\cite{dayabay12} and RENO~\cite{reno12} collaborations.  The fact 
that $\sin^2 2 \theta_{13}$ is now centered at around 0.1 means that the 
previously favored tribimaximal mixing pattern ($\sin^2 \theta_{23} = 1/2$, 
$\sin^2 \theta_{12} = 1/3$, $\theta_{13}=0$) is invalid, although the 
$A_4$ symmetry~\cite{mr01,m02,bmv03} used to obtain it~\cite{m04} is still 
applicable with some simple modifications~\cite{mw11,im12,mnr12}.  On the 
other hand,  in the simplest application~\cite{mr01,bmv03} of $A_4$, all 
the quark mixing angles are zero.  The question is whether there 
exists another symmetry which successfully yields both quark and lepton 
mass matrices, with good fits of all masses, mixing angles, and phases.  
The answer is yes, as elaborated below. 

Using the non-Abelian discrete symmetry $D_7$ of the heptagon, it has been 
shown~\cite{cm05} that the $CP$ violating phase of the quark mixing matrix 
may be predicted, whereas $D_7$ also yields a pattern~\cite{m05} for the 
neutrino mass matrix consistent with what is observed.  This pattern 
is previously derived using the symmetry $Q_8$~\cite{fkmt05}, and realizes 
a specific conjecture~\cite{fgm02} that the neutrino mass matrix has two 
texture zeros in the basis that charged-lepton masses are diagonal.

In Sec.~2 the symmetry $D_7$ is explained.  In Sec.~3 the assignments of 
quarks under $D_7$ are given with the accompanying Higgs structure and 
the resulting mass matrices.  In Sec.~4 numerical fits to the quark masses 
and mixing angles are given, with a prediction of the $CP$ violating phase. 
In Sec.~5 the assignments of leptons under $D_7$ are given with the 
accompaning Higgs structure and the resulting mass matrices.  In Sec.~6 
the neutrino mass matrix is analyzed to show that it allows for nonzero 
$\theta_{13}$ and a specific correlation between it and $\theta_{23}$ 
as well as $\delta_{CP}$.  Given that $\theta_{12}$ is close to the 
tribimaximal value, it prefers an inverted hierarchy of neutrino masses 
although a quasidegenerate pattern with either normal or inverted ordering 
cannot be ruled out.  In Sec.~7 there are some 
concluding remarks.

\section{Heptagonic Symmetry $D_7$}

The group $D_7$ is the symmetry group of the regular heptagon with 14 
elements, 5 equivalence classes, and 5 irreducible representations. 
Its character table is shown below.\\

\begin{table}[htb]
\begin{center}
\begin{tabular}{|c|c|c|c|c|c|c|c|}
\hline
class & $n$ & $h$ & $\chi_1$ & $\chi_2$ & $\chi_3$ & $\chi_4$ & $\chi_5$ \\
\hline
$C_1$ & 1 & 1 & 1 & 1 & 2 & 2 & 2 \\  
$C_2$ & 7 & 2 & --1 & 1 & 0 & 0 & 0 \\  
$C_3$ & 2 & 7 & 1 & 1 & $a_1$ & $a_2$ & $a_3$ \\  
$C_4$ & 2 & 7 & 1 & 1 & $a_2$ & $a_3$ & $a_1$ \\  
$C_5$ & 2 & 7 & 1 & 1 & $a_3$ & $a_1$ & $a_2$ \\  
\hline
\end{tabular}
\caption{Character Table of $D_7$.}
\end{center}
\end{table}

Here $n$ is the number of elements and $h$ is the order of each element. 
The numbers $a_k$ are given by $a_k = 2 \cos (2k\pi/7)$.  The character 
of each representation is its trace and must satisfy the following two 
orthogonality conditions:
\begin{equation}
\sum_{C_i} n_i \chi_{ai} \chi^*_{bi} = n \delta_{ab}, ~~~ 
\sum_{\chi_a} n_i \chi_{ai} \chi^*_{aj} = n \delta_{ij},
\end{equation}
where $n = \sum_i n_i$ is the total number of elements.  The number of 
irreducible representations must be equal to the number of eqivalence classes.

The three irreducible two-dimensional reprsentations of $D_7$ may be chosen 
as follows.  For {\bf 2}$_1$, let
\begin{eqnarray}
&& C_1~:~ \pmatrix{1 & 0 \cr 0 & 1}, ~~~ C_2~:~ \pmatrix{0 & \omega^k \cr 
\omega^{7-k} & 0}, ~(k=0,1,2,3,4,5,6), \nonumber \\ 
&& C_3~:~ \pmatrix{\omega & 0 \cr 0 & \omega^6}, \pmatrix{\omega^6 & 0 \cr 0 
& \omega}, ~~~  C_4~:~ \pmatrix{\omega^2 & 0 \cr 0 & \omega^5}, 
\pmatrix{\omega^5 & 0 \cr 0 & \omega^2}, \nonumber \\  
&& C_5~:~ \pmatrix{\omega^4 & 0 \cr 0 & \omega^3}, 
\pmatrix{\omega^3 & 0 \cr 0 & \omega^4},
\end{eqnarray}
where $\omega = \exp(2\pi i/7)$, then {\bf 2}$_{2,3}$ are simply obtained 
by the cyclic permutation of $C_{3,4,5}$. 

For $D_n$ with $n$ prime, there are $2n$ elements divided into $(n+3)/2$ 
eqivalence classes: $C_1$ contains just the identity, $C_2$ has the $n$ 
reflections, $C_k$ from $k=3$ to $(n+3)/2$ has 2 elements each of order $n$. 
There are 2 one-dimensional representations and $(n-1)/2$ two-dimensional 
ones.

The group multiplication rules of $D_7$ are:
\begin{eqnarray}
&& {\bf 1}' \times {\bf 1}' = {\bf 1}, ~~~ {\bf 1}' \times {\bf 2}_i = 
{\bf 2}_i, \\
&& {\bf 2}_i \times {\bf 2}_i = {\bf 1} + {\bf 1}' + {\bf 2}_{i+1}, ~~~ 
{\bf 2}_i \times {\bf 2}_{i+1} = {\bf 2}_i + {\bf 2}_{i+2}, 
\end{eqnarray}
where ${\bf 2}_{4,5}$ means ${\bf 2}_{1,2}$.  In particular, let 
$(a_1,a_2),(b_1,b_2) \sim {\bf 2}_1$, then
\begin{equation}
a_1b_2 + a_2b_1 \sim {\bf 1}, ~~~ a_1b_2-a_2b_1 \sim {\bf 1}', ~~~ 
(a_1b_1,a_2b_2) \sim {\bf 2}_2.
\end{equation}
In the decomposition of ${\bf 2}_1 \times {\bf 2}_2$, we have instead
\begin{equation}
(a_2b_1,a_1b_2) \sim {\bf 2}_1, ~~~ (a_2b_2,a_1b_1) \sim {\bf 2}_3.
\end{equation}

\section{Quark Sector}

We assign quarks as shown in Table 2 and 
Higgs doublets as shown in Table 3, together with an extra 
$Z_2^d \times Z_2^u$ symmetry.

\begin{table}[htb]
\begin{center}
\begin{tabular}{|c|c|c|c|c|c|c|}
\hline
symmetry & $[(u,d),(c,s)]$ & $(t,b)$ & $(d^c,s^c)$ & $b^c$ & $(u^c,c^c)$ & 
$t^c$ \\
\hline
$D_7$ & ${\bf 2}_1$ & ${\bf 1}$ & ${\bf 2}_1$ & ${\bf 1}$ & ${\bf 2}_2$ & 
${\bf 1}$ \\ 
\hline
$Z_2^d$ & + & + & -- & -- & + & + \\ 
\hline
$Z_2^u$ & + & + & + & + & -- & + \\ 
\hline
\end{tabular}
\caption{Quark assignments under $D_7 \times Z_2^d \times Z_2^u.$}
\end{center}
\end{table} 

\begin{table}[htb]
\begin{center}
\begin{tabular}{|c|c|c|c|c|c|}
\hline
symmetry & $\Phi_1$ & $\Phi_2$ & $\Phi_{3,4}$ & $\Phi_{5,6}$ & $\Phi_{7,8}$ \\
\hline
$D_7$ & ${\bf 1}$ & ${\bf 1}$ & ${\bf 2}_1$ & ${\bf 2}_2$ & ${\bf 2}_3$ \\ 
\hline
$Z_2^d$ & + & -- & -- & + & + \\ 
\hline
$Z_2^u$ & + & + & + & + & -- \\ 
\hline
\end{tabular}
\caption{Higgs doublet assignments under $D_7 \times Z_2^d \times Z_2^u.$}
\end{center}
\end{table} 

As a result, the $(u,c,t)$ mass matrix 
is diagonal, coming from the Yukawa terms $u u^c \phi_7^0 + c c^c \phi_8^0$ 
and $t t^c \phi_1^0$.  As for the $(d,s,b)$ mass matrix, the allowed Yukawa 
terms are $(d s^c + s d^c) \bar{\phi}_2^0$, $b b^c \bar{\phi}_2^0$, 
$b(d^c \bar{\phi}_4^0 + s^c \bar{\phi}_3^0)$, and $(d \bar{\phi}_4^0 + 
s \bar{\phi}_3^0)b^c$.  The resulting mass matrix is thus of the 
form~\cite{cm05}
\begin{equation}
{\cal M}_d = \pmatrix{0 & a & \xi b \cr a & 0 & b \cr \xi c & c & d},
\end{equation}
where $\xi = \langle \bar{\phi}_4^0 \rangle / \langle \bar{\phi}_3^0 \rangle$.

\section{Prediction of $CP$ Phase}

As in Ref.~\cite{cm05}, we can redefine the phases of ${\cal M}_d$ so that 
$a,b,c,d$ are real, but $\xi$ is complex.  Since ${\cal M}_u$ is diagonal, 
we have
\begin{equation}
V_L^\dagger {\cal M}_d V_R = \pmatrix{m_d & 0 & 0 \cr 0 & m_s & 0 \cr 0 & 0 & 
m_b},  ~~~ V_L^\dagger {\cal M}_d {\cal M}_d^\dagger V_L = \pmatrix{m_d^2 & 0 & 0 
\cr 0 & m_s^2 & 0 \cr 0 & 0 & m_b^2},
\end{equation}
where $V_L$ is the observed quark mixing matrix up to phase conventions. 
The structure of ${\cal M}_d{\cal M}_d^\dagger$ allows us to obtain the 
following first approximations:
\begin{eqnarray}
m_b \simeq \sqrt{c^2+d^2}, ~~~ V_{cb} \simeq {bd + \xi^* ac \over 
(1+|\xi|^2)c^2 + d^2}, ~~~ V_{ub} \simeq {ac + \xi bd \over c^2 + d^2},
\end{eqnarray}
where $a^2 << b^2$ and $|\xi|^2 << 1$ are assumed.  We now rotate 
${\cal M}_d{\cal M}_d^\dagger$ using
\begin{equation}
V_3 = \pmatrix{ 1 & 0 & V_{ub} \cr 0 & 1 & V_{cb} \cr -V_{ub}^* & -V_{cb}^* & 1}
\end{equation}
to obtain the $2 \times 2$ matrix
\begin{equation}
{\cal M}_2 {\cal M}_2^\dagger = \pmatrix{A & C \cr C^* & B},
\end{equation}
where
\begin{eqnarray}
A &=& a^2 + |\xi|^2 b^2 - |V_{ub}|^2 m_b^2, \\ 
B &=& a^2 + b^2 - |V_{cb}|^2 m_b^2, \\ 
C &=& \xi b^2 - V_{ub} V_{cb}^* m_b^2,
\end{eqnarray}
yielding
\begin{eqnarray}
m_s^2 &=& {1 \over 2} (B+A) + {1 \over 2} \sqrt{ (B-A)^2 + 4|C|^2}, \\ 
|V_{us}|^2 &=& {1 \over 2} - {1 \over 2}\sqrt{1-{4|C|^2 \over (B-A)^2 + 4|C|^2}},
\end{eqnarray}
where the phase of $V_{us}$ is that of $C$, and
\begin{equation}
m_d = |2abc\xi-a^2d|/m_s m_b.
\end{equation}
Using $|V_{us}| = 0.22534$, we find $|C|^2/(B-A)^2 = 0.05971$, and 
$m_s^2 >> m_d^2$ implies $A \simeq 0.05351B$, hence $m_s^2 \simeq 1.05349B$. 
Using these formulas, the 6 parameters $a,b,c,d,Re(\xi),Im(\xi)$ may then 
be determined and the $CP$ violating parameter $J$ is predicted.

\begin{figure}[htb]
\includegraphics[width=14cm]{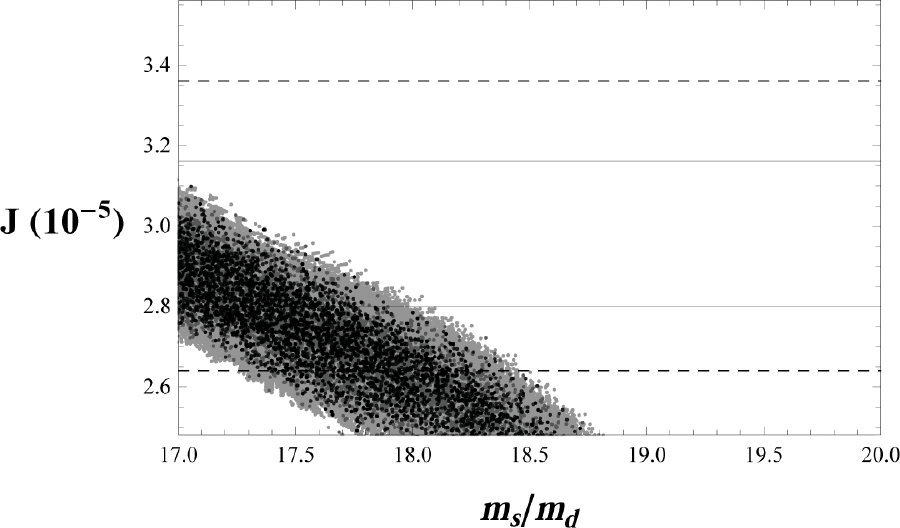}
\caption{The $CP$ violating parameter $J$ versus $m_s/m_d$. The solid (dash) 
lines indicate the one (two) standard-deviation bounds of $J$.}
\end{figure}

For our numerical analysis, we dispense with the approximate expressions 
and diagonalize ${\cal M}_d{\cal M}_d^\dagger$ directly.  We scan for 
solutions consistent with data on the 3 masses and 3 mixing angles, 
within one standard deviation of each parameter.  We then obtain 
$J$ numerically from the resulting $V_{CKM}$.  This is then the prediction 
of our model.  In Fig.~1 we plot $J$ versus $m_s/m_d$, which shows 
good agreement with data.  We use the 2008 updated values~\cite{xzz08} 
of $m_{d,s,b}$ evaluated at $M_W$:
\begin{eqnarray}
m_d(M_W) &=& 2.93~(+1.25/-1.21)~{\rm MeV}, ~~~\\ 
m_s(M_W) &=& 56 \pm 16~{\rm MeV}, ~~~ \\
m_b(M_W) &=& 2.92 \pm 0.09~{\rm GeV}, 
\end{eqnarray}
and the 2012 Particle Data Group (PDG)~\cite{pdg12} values of the mixing angles
\begin{eqnarray}
|V_{us}| &=& 0.22534 \pm 0.00065, ~~~\\ 
|V_{cb}| &=& 0.0412~(+0.0011/-0.0005), ~~~\\ 
|V_{ub}| &=& 0.00351~(+0.00015/-0.00014). 
\end{eqnarray}
Note that PDG also lists the condition 
$17< m_s/m_d < 22$ and the value of the $CP$ violating parameter is
\begin{equation}
J = 2.96~(+0.20/-0.16) \times 10^{-5}.
\end{equation}
We show in Table 4 sample values of $a,b,c,d,Re(\xi),Im(\xi)$ with 
the corresponding values of $m_d,m_s,m_b,|V_{us}|,|V_{ub}|,|V_{cb}|$ and 
$J$ as well as $m_s/m_d$.\\[10pt]
	
\begin{table}[htb]
\begin{center}	
\begin{tabular}{|c|c|c|c|c|c|c|}
\hline
$a$ (GeV)& $b$ (GeV) & $c$ (GeV)& $d$ (GeV) & $Re ( \xi) $ & 
$Im ( \xi)$ & $ m_s/m_d $ \\
\hline
$m_d$ (MeV) & $m_s$ (MeV)& $m_b$ (GeV) & $|V_{us}|$ &  $|V_{ub}|$ & $ |V_{cb}|$  
& J \\ 
\hline
\hline
0.0125 & 0.138 & 1.32 & -2.60 & 0.053 & -0.084 & 17.00 \\
\hline
3.89 & 66.2 & 2.92 & 0.22534 & 0.00355 & 0.0420 & $2.95 \times 10^{-5}$ \\
\hline
\hline
0.0124 & 0.139 & 1.34 & -2.60 & 0.058 & -0.084 & 17.25 \\
\hline
3.91 & 67.4 & 2.93 & 0.22532 & 0.00358 & 0.0420  & $2.89 \times 10^{-5}$\\
\hline
\hline
0.0123 & 0.138 & 1.40 & -2.60 & 0.064 & -0.087 & 17.50 \\
\hline
3.96 & 69.2 & 2.96 & 0.22519 & 0.00363 & 0.0409 & $2.76 \times 10^{-5}$\\
\hline
\hline
0.0122 & 0.138 & 1.39 & -2.55 & 0.068 & -0.084 & 17.75 \\
\hline
3.94 & 69.9 & 2.91 & 0.22501 & 0.00359 & 0.0415  & $2.70 \times 10^{-5}$\\
\hline
\end{tabular}
\caption{$D_7$ parameter fits of quark masses and mixing.}
\end{center}
\end{table}

\section{Lepton Sector}

Using again $D_7 \times Z_2^d \times Z_2^u$, we assign leptons as shown 
in Table 4 and Higgs triplets as shown in Table 5.
\begin{table}[htb]
\begin{center}
\begin{tabular}{|c|c|c|c|c|}
\hline
symmetry & $(\nu_e,e)$ & $[(\nu_\mu,\mu),(\nu_\tau,\tau)]$ & $e^c$ & 
$[(\mu^c,\tau^c)]$  \\
\hline
$D_7$ & ${\bf 1}$ & ${\bf 2}_1$ & ${\bf 1}$ & ${\bf 2}_3$ \\ 
\hline
$Z_2^d$ & + & + & + & + \\ 
\hline
$Z_2^u$ & + & + & + & + \\ 
\hline
\end{tabular}
\caption{Lepton assignments under $D_7 \times Z_2^d \times Z_2^u.$}
\end{center}
\end{table}

\newpage
\begin{table}[htb]
\begin{center}
\begin{tabular}{|c|c|c|}
\hline
symmetry & $\xi_1$ &  $\xi_{2,3}$  \\
\hline
$D_7$ & ${\bf 1}$ & ${\bf 2}_1$ \\ 
\hline
$Z_2^d$ & + & + \\ 
\hline
$Z_2^u$ & + & + \\ 
\hline
\end{tabular}
\caption{Higgs triplet assignments under $D_7 \times Z_2^d \times Z_2^u.$}
\end{center}
\end{table} 
As a result, the $(e,\mu,\tau)$ mass matrix 
is diagonal, coming from the Yukawa terms $e e^c \bar{\phi}_1^0$
and $\mu \mu^c \bar{\phi}_5^0 + \tau \tau^c \bar{\phi}_6^0$.  As for 
the Majorana $(\nu_e,\nu_\mu,\nu_\tau)$ mass matrix, the allowed Yukawa 
terms are $\nu_e \nu_e \xi_1^0$, $(\nu_\mu \nu_\tau + \nu_\tau \nu_\mu) 
\xi_1^0$, and $\nu_e (\nu_\mu \xi_3^0 + \nu_\tau \xi_2^0)$. 
The resulting mass matrix is thus of the form~\cite{m05}
\begin{equation}
{\cal M}_\nu = \pmatrix{a & c & d \cr c & 0 & b \cr d & b & 0},
\end{equation}
which was first derived using $Q_8$~\cite{fkmt05}, and realizes one of the 
conjectures of Ref.~\cite{fgm02}.

\section{Analysis of Neutrino Mass Matrix}

Rotating ${\cal M}_\nu$ to the tribimaximal basis using 
\begin{equation}
\pmatrix{\nu_1 \cr \nu_2 \cr \nu_3} = U_{TB}^\dagger \pmatrix{\nu_e \cr 
\nu_\mu \cr \nu_\tau} = \pmatrix{\sqrt{2/3} 
& -\sqrt{1/6} & -\sqrt{1/6} \cr \sqrt{1/3} & \sqrt{1/3} & \sqrt{1/3} 
\cr 0 & -\sqrt{1/2} & \sqrt{1/2}} \pmatrix{\nu_e \cr \nu_\mu \cr \nu_\tau},
\end{equation}
it becomes
\begin{equation}
{\cal M}_\nu^{(1,2,3)} = \pmatrix{m_1 & m_6 & m_4 \cr m_6 & m_2 & m_5 
\cr m_4 & m_5 & m_3},
\end{equation}
where
\begin{eqnarray}
m_1 &=& {1 \over 3}(2a + b - 2c - 2d), \\ 
m_2 &=& {1 \over 3}(a + 2b + 2c + 2d), \\
m_3 &=& -b, \\
m_4 &=& {1 \over \sqrt{3}}(-c+d),\\
m_5 &=& {1 \over \sqrt{6}}(-c+d) = {m_4 \over \sqrt{2}},\\
m_6 &=& {1 \over 3\sqrt{2}}(2a - 2b + c + d) = 
{1 \over 2 \sqrt{2}}(m_1+2m_2+3m_3).
\end{eqnarray}
If $m_4=m_5=m_6=0$, tribimaximal mixing is recovered.  In particular, 
$m_4 \neq 0$ or $m_5 \neq 0$ means that $\theta_{13} \neq 0$.  In previous 
studies, the special cases $m_4 \neq 0, m_5=m_6=0$~\cite{stw11,ckl12} and 
$m_5 \neq 0, m_4=m_6=0$~\cite{im12,mnr12,bmpsv12}  
have been explored.  The requirement from $D_7$ that $m_5 = m_4/\sqrt{2}$ 
is a new condition which will predict a special correlation between 
$\theta_{13}$ and $\theta_{23}$ as well as $\delta_{CP}$.  

Consider the unitary matrix $U_\epsilon$ such that
\begin{equation}
U^\dagger_\epsilon {\cal M}_\nu^{(1,2,3)}  ({\cal M}_\nu^{(1,2,3)})^\dagger U_\epsilon 
= \pmatrix{|m'_1|^2 & 0 & 0 \cr 0 & |m'_2|^2 & 0 \cr 0 & 0 & |m'_3|^2},
\end{equation}
then $U'_{\alpha i} = U_{TB} U_\epsilon$ is the lepton mixing matrix up to 
phases.  Let $U_\epsilon$ be approximately given by
\begin{equation}
U_\epsilon = \pmatrix{1 & \epsilon _{12} & \epsilon_{13} \cr \epsilon _{21} & 
1 & \epsilon_{23} \cr \epsilon _{31} & \epsilon_{32} & 1},
\end{equation}
then for $|m'_1|^2 \simeq |m_1|^2$, we have
\begin{equation}
\epsilon_{21} \simeq {-(m_6 m_1^* + m_2 m_6^*) \over |m_2|^2 - |m_1|^2}.
\end{equation}
In addition, since the effective neutrino mass $m_{ee}$ in neutrinoless double 
beta decay is given by
\begin{equation}
m_{ee} = |a| = |m_1 + m_2 + m_3|,
\end{equation}
whereas
\begin{equation}
m_3 = {1 \over 3} (2\sqrt{2} m_6 - m_1 - 2 m_2),
\end{equation}
we have the relationship
\begin{equation}
|m_3|^2 - m_{ee}^2 = {1 \over 3} (|m_2|^2 - |m_1|^2) [1 + 4 \sqrt{2} 
Re(\epsilon_{21})].
\end{equation}
Since $|m_2|^2 - |m_1|^2 \simeq \Delta m^2_{21}$ is very small, this model 
predicts $m_{ee} = |m_3|$ to a very good approximation.  The structure of 
Eq.~(38) also shows that an inverted ordering of neutrino masses is 
expected, although the quasidegenerate limit is also possible in which 
case either inverted or normal ordering may occur.  In the following we 
focus on the inverted case, i.e. $|m_3| < |m_1| < |m_2|$.

For $m_4 \neq 0$, $\nu_3$ is rotated to $\nu'_3$ according to
\begin{eqnarray}
\epsilon_{13} \simeq {m_1 m_4^* + m_4 m_3^* \over |m_3|^2 - |m_1|^2},  ~~~
\epsilon_{23} \simeq {m_2 m_4^* + m_4 m_3^* \over \sqrt{2} (|m_3|^2 - |m_1|^2)}.
\end{eqnarray}
As a result,
\begin{eqnarray}
U'_{e3} &\simeq& \sqrt{2 \over 3} \epsilon_{13} + \sqrt{1 \over 3} \epsilon_{23} 
\simeq  {-m_4(m_1+2m_2)^* + m_4^* (2m_1+m_2) 
\over \sqrt{6} (|m_1|^2 - |m_3|^2)}, \\ 
U'_{\mu3} &\simeq& -{1 \over \sqrt{2}} - {1 \over \sqrt{6}} \epsilon_{13} + 
{1 \over \sqrt{3}} \epsilon_{23} \simeq  -{1 \over \sqrt{2}} - 
 {(m_1-m_2) m_4^* \over \sqrt{6} (|m_1|^2 - |m_3|^2)}, \\ 
U'_{\tau3} &\simeq& {1 \over \sqrt{2}} - {1 \over \sqrt{6}} \epsilon_{13} + 
{1 \over \sqrt{3}} \epsilon_{23} \simeq  {1 \over \sqrt{2}} - 
{(m_1-m_2) m_4^* \over \sqrt{6} (|m_1|^2 - |m_3|^2)},
\end{eqnarray}
If all parameters are real, then for $U'_{e3} = 0.16$,  $\sin^2 2 \theta_{23}$ 
would be 0.80, which is ruled out by present data, i.e. $\sin^2 2 \theta_{23} 
> 0.92$.  However, a fit may be obtained for complex values.

We go back to Eq.~(25) and observe that $a,c,d$ may be chosen real, so only 
$b$ is complex.  This means that $m_4$ is real as well as $2m_1-m_2$, and 
for $m_6 = 0$, $m_3 = -(m_1+2m_2)/3$.  Writing $m_{1,2}$ as $m_{1,2}~e^{i\phi_{1,2}}$ 
with $m_2 \simeq m_1$ and $\sin \phi_2 = 2 \sin \phi_1$, we obtain
\begin{eqnarray}
U'_{e3} &\simeq& {m_1 m_4 \over \sqrt{6} \Delta m^2_{32}} [-\cos \phi_1 + 
\cos \phi_2  - 9i \sin \phi_1], \\ 
U'_{\mu3} &\simeq& - {1 \over \sqrt{2}} + {m_1 m_4 \over \sqrt{6} \Delta m^2_{32}} 
[\cos \phi_1 - \cos \phi_2 - i \sin \phi_1], \\ 
U'_{\tau3} &\simeq& {1 \over \sqrt{2}} + {m_1 m_4 \over \sqrt{6} \Delta m^2_{32}} 
[\cos \phi_1 - \cos \phi_2 - i \sin \phi_1],
\end{eqnarray}
where $\cos \phi_2 = \pm \sqrt{1-4\sin^2 \phi_1}$.  We then have 
\begin{equation}
\sin^2 \theta_{13} = {|U'_{e3}|^2 \over 1 + |\epsilon_{13}|^2 + 
|\epsilon_{23}|^2}, ~~~ \tan^2 \theta_{23} = {|U'_{\mu 3}|^2 \over |U'_{\tau 3}|^2}.
\end{equation}
Since
\begin{equation}
|m_3| \simeq {\sqrt{\Delta m^2_{32}} \sqrt{5 + 4 \cos (\phi_2 - \phi_1)} \over 
2 \sqrt{1 - \cos (\phi_2 - \phi_1)}},
\end{equation}
the above equations relate $|m_3| = m_{ee}$ with $\theta_{13}$ and $\theta_{23}$. 
If we fix $\theta_{13}$, we then obtain $|m_3|$ as a function of $\theta_{23}$.
We plot in Fig.~2 our model predictions for $|m_{1,2}|$ and $|m_3|=m_{ee}$ versus 
$\sin^2 2 \theta_{23}$.  The other data points are taken to be their 
experimental central values.
\begin{figure}[htb]
\includegraphics[width=14cm]{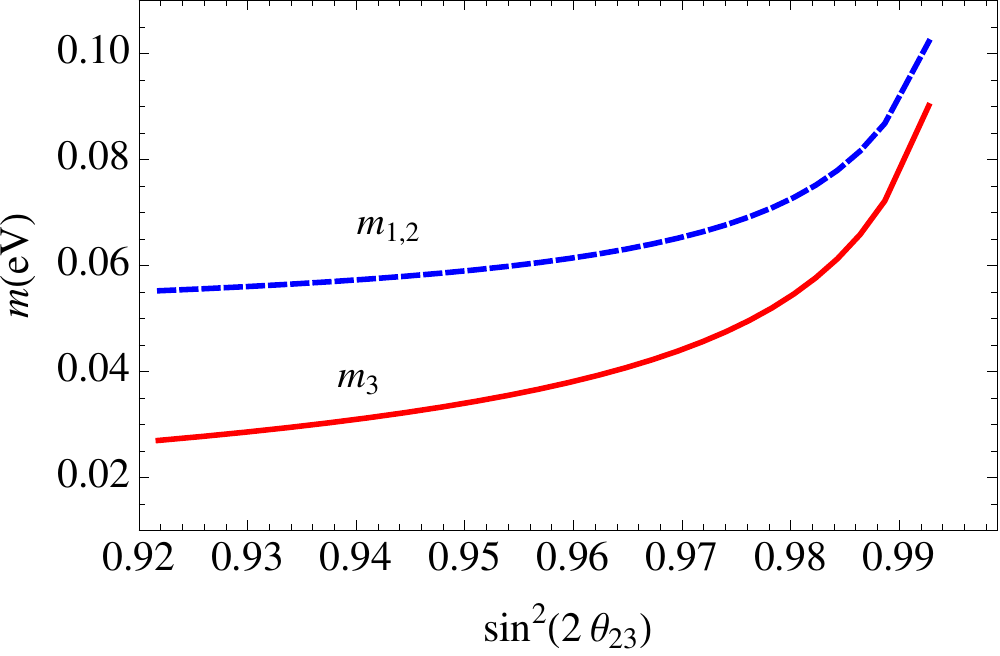}
\caption{Neutrino masses $m_{1,2}$ and $m_3=m_{ee}$ versus 
$\sin^2 2 \theta_{23}$.}
\end{figure}

If we rotate ${\cal M}_\nu^{1,2,3}({\cal M}_\nu^{1,2,3})^\dagger$ by
\begin{equation}
U'_\epsilon = \pmatrix{1 & 0 & \epsilon_{13} \cr 0 & 
1 & \epsilon_{23} \cr -\epsilon _{13}^* & -\epsilon_{23}^* & 1},\\[10pt]
\end{equation}
we obtain the $2 \times 2$ mass-squared matrix spanning $\nu'_{1,2}$. 
This differs from the $2 \times 2$ submatrix in the tribimaximal 
basis by terms quadratic in $m_4$ which are important in obtaining the 
correct $\Delta m^2_{21}$ and Eq.~(36) becomes modified.  However, we 
can adjust $|m_2|$ versus $|m_1|$ as well as $m_6$ to fit the data. 
These adjustments will have negligible effects on $|m_3|$. 

\begin{figure}[htb]
\includegraphics[width=14cm]{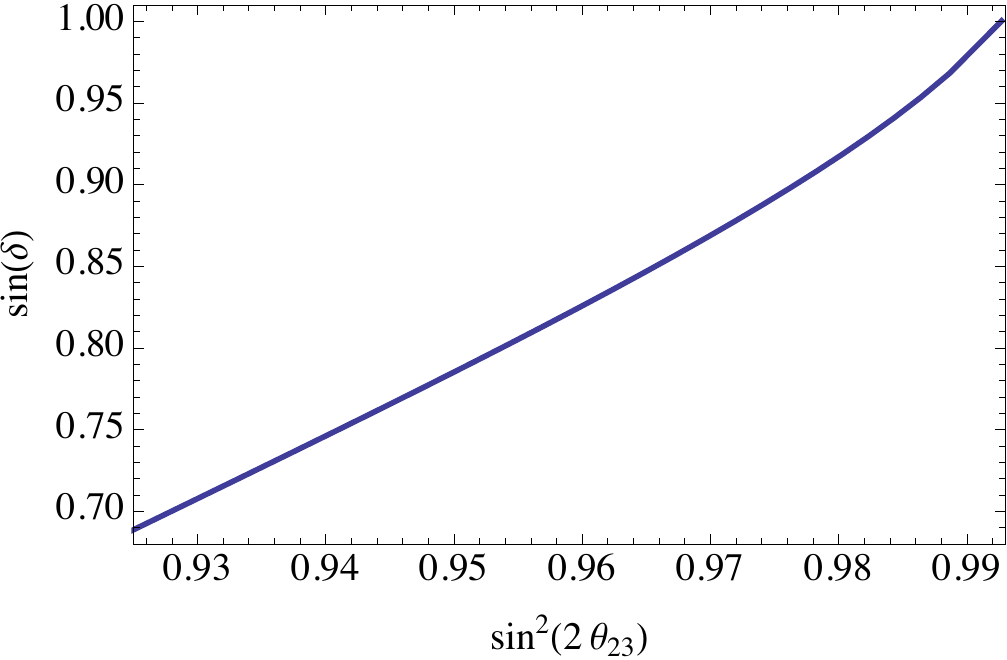}
\caption{The $CP$ violating parameter $|\sin \delta_{CP}|$ versus 
$\sin^2 2 \theta_{23}$.}
\end{figure}
We plot in Fig.~3 our model prediction for $|\sin \delta_{CP}|$ versus 
$\sin^2 2 \theta_{23}$.  To obtain $\sin \delta_{CP}$, we use
\begin{equation}
U'_{e2} \simeq {1 \over \sqrt{3}}, ~~~ U'_{\mu 2} \simeq  {1 \over \sqrt{3}} 
+ {1 \over \sqrt{2}} \epsilon_{23}^*, ~~~ J = Im (U'_{e2} U'_{\mu 3} 
{U'_{\mu 2}}^* {U'_{e3}}^*),
\end{equation}
from which we find (using $U'_{\mu 2} = |U'_{\mu 2}| e^{i \theta_{\mu 2}}$, etc.)
\begin{equation}
\sqrt{2 \over 3} \cos \theta_{23} \sin \delta \simeq 
|U'_{\mu2}| \sin (\theta_{\mu 3} - \theta_{\mu 2} - \theta_{e 3}).
\end{equation}

\section{Concluding Remarks}

We have studied a specific pattern for both quark and lepton mass matrices. 
In both cases, one mass matrix is diagonal (${\cal M}_u$ and ${\cal M}_e$), 
whereas the other has two zeros (${\cal M}_d$ and ${\cal M}_\nu$). 
In the case of ${\cal M}_\nu$, the assumption that it is Majorana 
corresponds to one of the conjectures of Ref.~\cite{fgm02}, whereas 
the Dirac mass matrix ${\cal M}_d$ requires further restrictions 
to make it predictive, as first proposed in Ref.~\cite{cm05} using 
the non-Abelian discrete symmetry $D_7$.  The conjectured form of 
${\cal M}_\nu$ was first derived~\cite{fkmt05} using $Q_8$, but it 
may also be obtained~\cite{m05} using $D_5$ or $D_7$.  Here we consider 
$D_7$ as the unifying symmetry for both quarks and leptons.

The $CP$ violating parameter $J$ in the quark sector is constrained in this 
model by $m_d,m_s,m_b,|V_{us}|,|V_{ub}|,|V_{cb}|$.  Within one standard 
deviation of all six measurements, we obtain $J$ in agreement with data.  
In the neutrino sector, we obtain $|m_{1,2}|$ as well as $|m_3| = m_{ee}$ as 
functions of $\sin^2 2 \theta_{23}$ and also predict $\sin \delta_{CP}$ as 
a function of $\sin^2 2 \theta_{23}$.
 
\section*{Acknowledgment} 
We thank M. Frigerio and A. Villanova del Moral for important 
communications. 
This work is supported in part by the U.~S.~Department of Energy under 
Grant No.~DE-AC02-06CH11357.

\bibliographystyle{unsrt}

\end{document}